%% file: paper108.tex
\documentclass[manuscript,screen]{acmart}
\usepackage{mathtools}

\AtBeginDocument{%
  }

\setcopyright{none}
\copyrightyear{2022}
\acmYear{2022}
\acmDOI{XXXXXXX.XXXXXXX}

\acmISBN{978-1-4503-XXXX-X/18/06}

\newcommand{\codeName}[1]{\textsc{#1}}
\newcommand{\keyword}[1]{\texttt{#1}}

\definecolor{myred}{HTML}{FF4136}
\newcommand{\todo}[1]{\textcolor{myred}{FIXME: #1}}
\begin{document}

\title{Gene Expression Programming for Quantum Computing}

\author{Gonzalo Alvarez}
\affiliation{%
	\institution{Oak Ridge National Laboratory}
	\city{Oak Ridge}
	\country{USA}}
\orcid{1234-5678-9012}

\author{Ryan Bennink}
\affiliation{%
	\institution{Oak Ridge National Laboratory}
	\city{Oak Ridge}
	\country{USA}}
\orcid{1234-5678-9013}

\author{Stephan Irle}
\affiliation{%
	\institution{Oak Ridge National Laboratory}
	\city{Oak Ridge}
	\country{USA}}
\orcid{1234-5678-9014}

\author{Jacek Jakowski}
\affiliation{%
	\institution{Oak Ridge National Laboratory}
	\city{Oak Ridge}
	\country{USA}}
\orcid{1234-5678-9015}


\begin{abstract}
We introduce \codeName{QuantumGEP}, a scientific computer program that uses gene expression programming (GEP) 
to find a quantum circuit that either (i) 
maps a given set of input states to a given set of output states, or (ii) transforms a fixed initial state to minimize a given physical quantity of the output state. \codeName{QuantumGEP} is a driver program that uses \codeName{evendim}, a generic computational engine for GEP,
both of which are free and open source. We apply \codeName{QuantumGEP} as a powerful solver for MaxCut in graphs, and for
condensed matter quantum many-body Hamiltonians. 
\end{abstract}



\keywords{genetic algorithms, gene expression programming, quantum computing, condensed matter, quantum chemistry}
\maketitle

\section{Introduction}One of the fundamental problems in quantum computing is to find a quantum circuit
that, when applied to a prepared initial state, yields the ground state of a quantum many-body Hamiltonian.
This is a conventional yet difficult problem that can in principle be solved by machine learning. 
We have chosen gene expression programming (GEP) \cite{re:ferreira01, re:ferreira06} as a tool
to generate quantum circuits for applications.
Only recently has interest in automatically generated quantum circuits increased \cite{PhysRevA.105.052414}, so that
our goal of introducing a free and open source tool for quantum circuit generation appears timely.

Other solutions to the same or similar problems have used
the quantum approximate optimization algorithm (QAOA),
\cite{https://doi.org/10.48550/arxiv.1411.4028,https://doi.org/10.48550/arxiv.1602.07674,Shaydulin_2019} a variational 
quantum-classical algorithm for combinatorial optimization problems. 
An iterative version of
QAOA  has recently been developed\cite{https://doi.org/10.48550/arxiv.2005.10258} so that it is problem-tailored;
it can also be adapted to specific hardware constraints.

In our approach using GEP, a set of initially random quantum circuits (``population'') is evolved by repeatedly introducing variations of the existing circuits (``mutations'') and retaining only those individuals that perform best at the prescribed computational task (have the highest ``fitness''). After sufficiently many iterations (``generations''), the population should consist mostly of circuits that perform well at the prescribed task.
To accomplish the mutations, each individual is specified by a string of symbols called a genome. Mutations are small random changes to these strings of symbols, yielding generally new and different circuits. The fitness of an individual is evaluated via simulation, but in principle could be evaluated efficiently by executing the circuit on a quantum computer.

This work introduces \codeName{QuantumGEP}, a scientific computer program that uses GEP to find a quantum circuit that either (i) 
maps a given set of input states to a given set of output states, or (ii) transforms a fixed initial state to minimize a given physical quantity of the output state.
Both tasks are common in quantum mechanics.
Training a quantum computer to learn a polynomial \cite{re:zhou01} or perform a quantum Fourier transform are examples of the first kind of task.
Finding the ground state energy of a molecule \cite{re:mccaskey19} or solving a combinatoric optimization problems \cite{Lotshaw_2021,mermin2007quantum} are examples of the second kind of task.
\codeName{QuantumGEP} is a driver program that uses \codeName{evendim}, a generic computational engine for GEP
that we developed from scratch; both \codeName{QuantumGEP} and \codeName{evendim} are free and open source.
Section~\ref{sec:theory} discusses the representation of quantum circuits by GEP strings, and the two algorithms used,
one for arbitrary function representation, and one for ground state of a Hamiltonian.
We discuss the GEP features currently available in \ref{sec:codeStructure}. Section~\ref{sec:caseStudies} discusses two
applications, one to the MaxCut problem in graphs, and one to ground state properties of many-body Hamiltonians as they
appear in condensed matter and quantum chemistry. Section~\ref{sec:summary} presents a summary and outlook for future work.
\section{Theory and Implementation}\label{sec:theory}

\subsection{Gene Expression Programming}

\begin{figure}
	\includegraphics[width=0.45\textwidth]{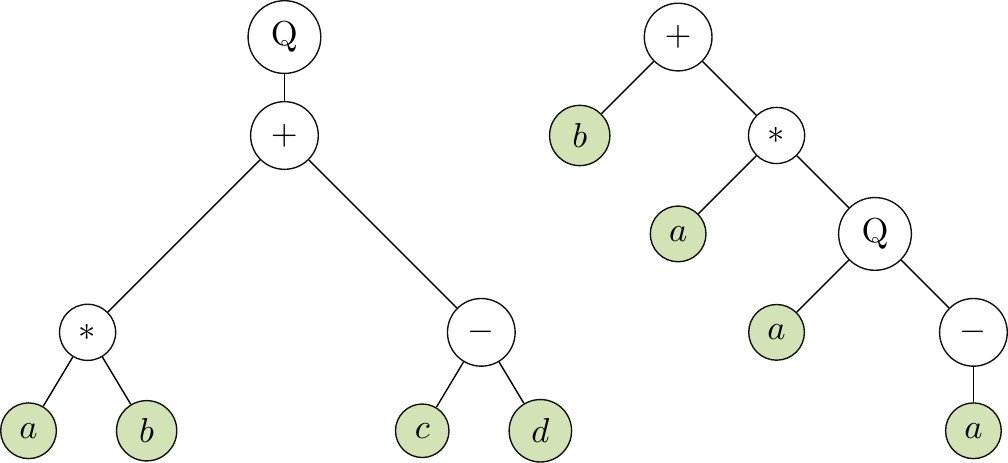}
	\caption{\label{fig:codingAndNot}(a) AST for $\sqrt{a*b+(c-d)}$, which has the
		string representation Q+*-abcd, operator (or primitives in GEP) in $\{$Q$,+,-\}$ and 
		leaves (shown shaded in the figure) in $\{a,\,b,\,c,\,d\}$. 
		(b) The set of operators here is as in
		(a), with inputs $a$ and $b.$ The twenty character
		AST string +b*aQa-ababbabbbabab has a coding region
		(the first eight characters)
		that is shown pictorially as
		an AST. The last twelve characters are non coding \cite{re:ferreira06}.}
\end{figure}

Any computer program \cite{re:arora07} can be written in abstract syntax tree (AST) form; Figure \ref{fig:codingAndNot}(a)
shows an example. To understand this figure, one first needs to consider what primitives or operators to
include, which depends on the problem. In the example, we have chosen from the set $\{Q, *, +, -\},$ where $Q$
is the square-root operator, and the other symbols stand for their usual algebraic meaning. The operator or
primitive $Q$ is unary, that is, it takes only one argument whereas the other operators are binary requiring two
arguments each. In the example, the leaves, or inputs, are taken from the set $\{a, b, c, d\}$ and can be
considered inputs or numbers. We can represent any AST not just pictorially, but also in string form: we do
this by writing down the primitives and leaves in breadth-first order. The AST shown in Figure \ref{fig:codingAndNot}(a) has thus
the string representation $Q+*-abcd.$
In the process of our artificial intelligence software finding a suitable mathematical function, we can think of an AST as
an individual life form; many ASTs form a population of life forms. More specifically, we think of the AST
as the genome and body of an individual; when we have a population of genomes they reproduce and create
a new generation or population from the previous one. This is done by combining and mutating genes, and selecting the survivors that
come closest to the unknown function.

Every individual genome or AST can be assigned a fitness
number indicating how close the ASTs outputs are to those of the
unknown function for each input. The unknown function depends
on the problem: it is only known through a series of inputs and
outputs. When the genetic simulation converges, the ``last''
population of individual ASTs will all constitute accurate ASTs
(that is, computer programs) that output the function our problem
required. We could then select the shortest of these ASTs, or one
that shows a previously unidentified pattern.

The overarching difficulty with the just described approach lies
in the combinations and mutations: they do not necessarily create a
new valid tree, that is, the resulting AST may not be syntactically
correct. A syntactically incorrect tree cannot have a fitness value
because it cannot be evaluated, and the approach just described fails. 
Of course, one could restrict the mutations in complicated
ways, but this would have a large and detrimental impact on convergence.
That is why non-coding regions are used \cite{re:ferreira06}
in the AST, when represented as a string.
Figure~\ref{fig:codingAndNot}(b) shows how a fixed length string of thirty characters has a
coding region of seven characters, and a non-coding region of
twenty-three characters. By introducing non-coding regions, all
mutations become valid and are allowed, restricting mutations is
no longer necessary, and convergence vastly improves. Obviously, the structural organization of genes
must be preserved, which is achieved by always maintaining the boundaries between coding and non-coding
regions and by not allowing symbols from the function set on the non-coding region. The resulting
coding region can be deduced automatically for any string, so that the genotype has been decoupled from
the phenotype.

With this background in place, we can now list many features of gene expression programming (GEP):
multiple outputs by having multiple genes to form a ``chromosome,'' and multiple chromosomes to form a
``cell;'' numeric constants can be added and evolved; blocks of ``genetic'' material can be used in
automatically defined functions that aid problem solving;
genes can be subjected to large variety of mutations and recombinations.

\subsection{Representation of Circuits}

\begin{figure}[h]\centering{%
		\includegraphics[width=0.45\textwidth]{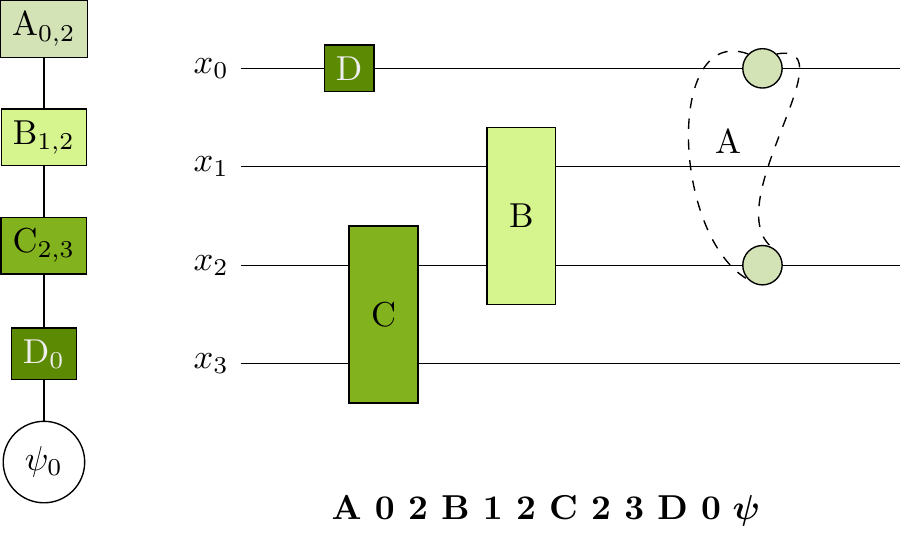}
		\caption{\label{fig:circuit}
			An example of a 4-bit quantum circuit with gates A, B, C, and D in order
			to describe the circuit encoding in gene expression programming.
			Representation of the previous figure's quantum circuit in GEP.
			The ``primitives'' or gates are D$_{0}$, C$_{2, 3}$, B$_{1, 2}$, and A$_{0, 2}$.
			Every primitive takes one input and produces one output.
			The final output is A$_{0,2}$B$_{1,2}$C$_{2,3}$D$_0\psi_0$.}}
\end{figure}

To represent the four bit quantum circuit of Fig.~\ref{fig:circuit} in gene expression programming,
we consider the Hilbert space $\mathcal{H}=\mathbb{C}^{2^N}$, with its usual inner product,
so that the inputs are in $\mathcal{H}$.
Gates are the GEP primitives, and are functions of $\mathcal{H}$ into $\mathcal{H}$.
One-bit gates act only on one of the $N$ bits; there are $N$ of them for each type (like $N$ Hadamard gates),
and the bit they act on is indicated with a subscript.
The multiplicity of each type of two-bit gate (like the CNOT gate) is $N(N-1)$;
the multiplicity of each type of three-bit gate (like the CCNOT gate) would be $N^3-3N(N-1)-N$.
The gene for Fig.~\ref{fig:circuit} is thus encoded by the string A$_{0,2}$B$_{1,2}$C$_{2,3}$D$_0\psi_0$.

\subsection{Optimization Procedure}\label{sec:optimization}
\codeName{QuantumGEP} solves two types of problems with two slightly different algorithm variants.
The first problem is finding a quantum circuit $\mathcal{C}_{\textrm{unknown}}$ that 
implements a quantum function given by a set of input-output pairs $
|\psi_{\textrm{in}}^{i}\rangle \to  |\psi_{\textrm{out}}^{i}\rangle$ for $i=1,2,\ldots,D$.
These input-output pairs constitute training data for the quantum function to be learned.
The second problem is finding a quantum circuit
$\mathcal{C}$ that, when applied to a fixed initial state $|\psi_\textrm{in}\rangle$, minimizes the expectation value of some physical observable $H$.
An example of this second type of problem is finding the ground state of a many-body Hamiltonian.

The gene expression programming procedure to solve either of these two problems is as follows. (For simplicity, the description is written as though for problem 1 there is a single input-output relation $|\psi_\textrm{in}\rangle \to |\psi_\textrm{out}\rangle$ to be satisfied. In the case of multiple input-output pairs, steps 4 and 5 should be modified to compute the average fitness of each circuit over all input-output pairs.)

\begin{enumerate}
	\item Start with an original population of $M$ circuits generated randomly.
	\item By gene expression programming mutations and combinations of the 
	existing size $M$ population, generate $M'=2M$ new circuits
	$\{\mathcal{C}(\varphi)\}_{0\le j<M'}$. These circuits may depend on
	$K$ continuous variables
	$\varphi\equiv\{\varphi_k\}_{0\le k<K}$. To understand why this is so, consider that the rotation gate
	may depend on the angle of rotation, and, in general, gates may depend on arbitrary parameters that
	we collectively call $\varphi$.
	\item For every circuit $j$, define the $\varphi-$dependent state $|\psi_j(\varphi)\rangle \equiv \mathcal{C}_j(\varphi)|\psi_\textrm{in}\rangle$.
	\item For every circuit $j$, compute the pre-fitness function $P_j$: 
	\begin{equation}
		P_j(\varphi) \equiv
		\begin{cases*}
			|\langle \psi_{\textrm{out}} | \psi_j(\varphi)\rangle|^2\, & for problem 1 \\
			-\langle \psi_j(\varphi)| H | \psi_j(\varphi)\rangle\,  & for problem 2
		\end{cases*}\label{eq:fitness}
	\end{equation}
	and find the $\varphi$ where the maximum occurs; let's call it $\varphi_{\textrm{max}}$.
	\item For every circuit $j$, calculate its fitness $F_j \equiv P_j(\varphi_{\textrm{max}})$.
	\item Eliminate the $M$ circuits with least fitness and keep the remaing $M$ circuits as the source for the next generation.
	\item Go to step 3. 
\end{enumerate}
For the studies described in this paper, $M=100$.

\subsection{Code Structure}\label{sec:codeStructure}
The free and open source code can be found at \nolinkurl{https://code.ornl.gov/gonzalo_3/evendim} and at
\nolinkurl{https://github.com/g1257/evendim}.
\codeName{QuantumGEP} can be regarded as ``artificial intelligence (AI)'' to distinguish it from the generated ASTs that run inside it: the
AI can then be considered also as a virtual machine that runs ASTs in order to calculate their fitness. \codeName{QuantumGEP}
can be divided into two coding sections: a computational engine that is independent of fitness function and the set of primitives and inputs (\codeName{evendim});
and a problem-dependent part: a fitness function and a set of primitives and nodes that can be programmed for each
particular problem to be solved (our quantum circuit
problem solver \codeName{QuantumGEP}). An arithmetic set has been implemented to test compilation on different
platforms, and as a fast way to test development ideas.

The two types of algorithms discussed in subsection \ref{sec:optimization} create two code paths, distinguished only by C++ templates; both
paths are instantiated in the same executable and are chosen at runtime from the input file, by setting RunType to either
FunctionFit or GroundState. \codeName{QuantumGEP} reads from the input file all non problem-specific parameters, such as number of
generations and population. A parameters object is created, as well as an evolution object, and from both an engine object
is instantiated with an initial random population;
this last object calls its member function evolve() in a loop as many times as the number of generations, with
an optional early stop which is useful in cases where the maximum fitness value is known.

The evolve() function in the computational engine starts by creating new individuals, in its first call from the initial population, and in
subsequent calls from the surviving individuals. It does so by using the following four algorithms in succession:
(1) one-point recombination, (2) two-point recombination, (3) mutation, (4) inversion, and (4) swap; all these
algorithms were implemented as detailed in \cite{re:ferreira06}, and we briefly review them in the following.

Recombination involves two parent chromosomes and results in two new individuals.
One-point recombination consists of paring the parent chromosomes side by side,
choosing a random point at which the parent chromosomes are split up, and exchanging
the genetic content after the recombination point between the two chromosomes.
Two-point recombination pairs the chromosomes side by side as before, chooses two random points,
and exchanges the genetic material between these two points, creating two new individuals.
A mutation changes one character of the string representation of the chromosome; in the head any character
can change to any other, so any function can be changed to any other without regards to the number of arguments.
In the tail, terminal or leafs are changed only into terminal or leafs so that the head and tail structure of the
chromosome is preserved by the mutation. Inversion involves inverting the characters in the head of the chromosome,
and does not affect the tail. \codeName{evendim} inverts the complete head even though subsets of the head could be inverted instead.
Finally, a swap exchanges two characters in the string representation of a chromosome such that the head and tail
structure is preserved.

After the new population has been created, which also includes the surviving individuals from the previous generation,
an optional canonicalization procedure is applied. For quantum circuits, the canonicalization orders the gates by the
bit or bits they acts on; there is also here an opportunity for symbolic simplifications: for example, the Pauli matrix gate 
$\sigma^z$ if applied on the same bit twice yields the identity. More complicated simplifications could be added here as well,
based on commutation rules among operators or gates. Finally, \codeName{QuantumGEP} sorts the individuals by fitness, its definition
depending on the problem to be solved as shown in Eq.~(\ref{eq:fitness}), and discards as many individuals with lowest fitness
as needed to obtain the population supplied in the input file.

\section{Case Studies} \label{sec:caseStudies}
\subsection{Graph Maximum Cut}
Finding the maximal cut of a graph is a well known problem (the MaxCut problem) in quantum computing 
\cite{Lotshaw_2021,farhi2014quantum,wang2018quantum,zhou2020quantum,MaxCutRequiresHundredsQubits,Love2020Bounds,re:mcelroy05}
The maximal cut is a partition of a graph's vertices into disjoint sets $A$ and $B$, 
such that the number of edges between the set $A$ and the set $B$ is the largest.

MaxCut for a graph $G$ is equivalent to finding the ground state of the
classical Ising model \cite{RevModPhys.39.883} with Hamiltonian $H=\sum_{i,j\in E(G)} S_i S_j$, where
the sum is over vertices (or ``sites'') that form edges in $G$, and $S_i\in\{-1, 1\}$ is the ``spin'' configuration
at vertex $i$, such as a radical electron embedded in an atomic lattice. \codeName{QuantumGEP} used with this Hamiltonian excels at solving MaxCut, by which 
we mean that it excels at finding a quantum circuit that yields the ground state of the graph Ising model from an initial state.	
We discuss in detail two examples.

\begin{figure}[h]\centering{%
		\includegraphics{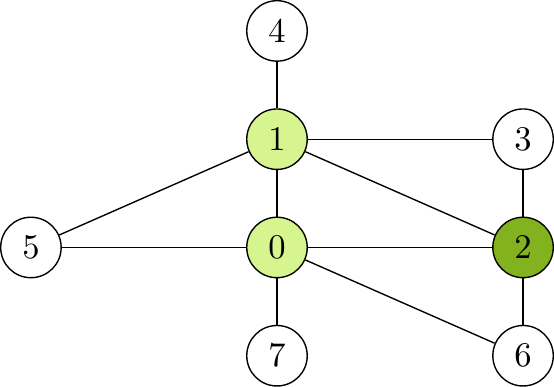}}
	\caption{\label{fig:graph34} Graph 34. A solution is given by 3 or 7. The first number (3) is 00000011 in binary, which means
		sites 0, 1, have negative signs and form the maximal cut; The other solution found is 7, 00000111 in binary, which
		corresponds to sites 0, 1, and 2 having negative signs and forming the maximal cut. 
		Vertices 0, 1, and 2 have been shaded, with vertex 2 shaded more to indicate that it appears only in the second solution.
		In all cases energy equals -5, fitness 5.}
\end{figure}

Figure~\ref{fig:graph34} shows our first example graph; it has 8 vertices.
After it converges we see the following circuit:
Ry$_0$:$\pi$ Ry$_1$:$\pi$ Ry$_2$:$\phi_0$ P$_4$ P$_4$ 
with fitness $4.9999(6)$.
We know this circuit is a perfect solution, that is, it is a true ground state because
(i) its fitness is $5$ within numerical error, and the true ground state of this graph is
$-5$. The maximal cut is found by looking at the basis states with weight
greater than a certain threshold $\epsilon$.
For numerical stability we use $\epsilon=1e-4$ but the actual value is not crucial.
With this criterion, \codeName{QuantumGEP} yields the basis states 3 and 7, that is, two basis states.
State 3 is binary 00000011, which means that sites 0 and 1
have negative ``spin'' and the rest positive ``spin.''
Similarly, the other solution is 7 (binary  00000111),
which means that sites 0, 1, and 2 have negative ``spin'' and the rest positive ``spin.''

An initial state needs to be provided, representing a single graph or a superposition of graphs.
We have chosen the graph without a cut, that is 00000000, which can also be
described as a configuration with all ``spins'' positive. We have found that the solutions are
robust to this initial condition. \codeName{QuantumGEP} allows the specification of the initial step in the input file.

\begin{figure}[h]\centering{%
		\includegraphics{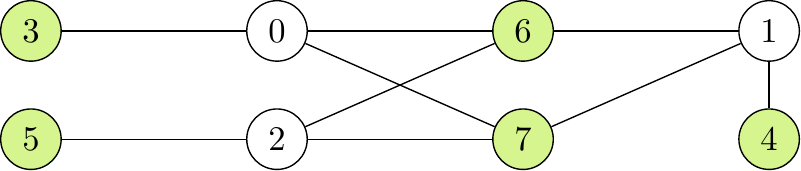}}
	\caption{\label{fig:graph35} Graph 35, which is number 4648 in the qaoa paper \cite{Lotshaw_2021}. A solution is given by 248, which in binary is
		11111000, which means
		sites 3 to 7 have negative signs and form the maximal cut; energy equals -9, fitness 9.}
\end{figure}

Figure~\ref{fig:graph35} shows our second example labeled 4648 in \cite{Lotshaw_2021}.
After it converges we obtain the quantum circuit
Ry$_3$:$\pi$ P$_3$ P$_3$ Ry$_4$:$\pi$ Ry$_5$:$\pi$ Ry$_6$:$\pi$ Ry$_7$:$\pi$
with fitness $8.9999(9)$  and weight only on basis state 248.
Following a similar reasoning than before, we conclude that this is a perfect solution circuit,
and 248 (binary 1111000), indicates that vertices 3 to 7 have
negative signs and form the maximal cut for this graph. 

\subsection{Condensed Matter Models}

We apply \codeName{QuantumGEP} to the XX model on a linear chain of sites; the model is given by
$H = J_x\sum_{i} \sigma^x_i \sigma^x_{i + 1}$, where $J_x$ is a real number acting as coupling constant, $\sigma^x$ is the Pauli $x$ matrix,
and the usual notation for tensor product of operators has been used.
We aim at finding the quantum circuit that produces the ground state of this model from an initially
prepared state.
We choose the vector $|0000\rangle$, periodic or open boundary conditions as detailed below,
coupling $J_x= 1$, and quantum gates Ry and P.
All these parameters can be chosen without recompilation from the input file.

After a few generations we get various perfect ``individuals,'' that is, quantum circuits. For example,
with periodic boundary conditions we get
Ry$_0$:$3\pi/2$ Ry$_1$:$\pi/2$ Ry$_2$:$3\pi/2$ Ry$_3$:$\pi/2$ with fitness 3.9971(6) 		
which yields the exact answer for the ground state of four sites, and another state can be obtained by symmetry.
With open boundary conditions, we obtain the same vector, but fitness (or minus the energy) equals 3.

\begin{figure}
	\includegraphics[width=0.45\textwidth]{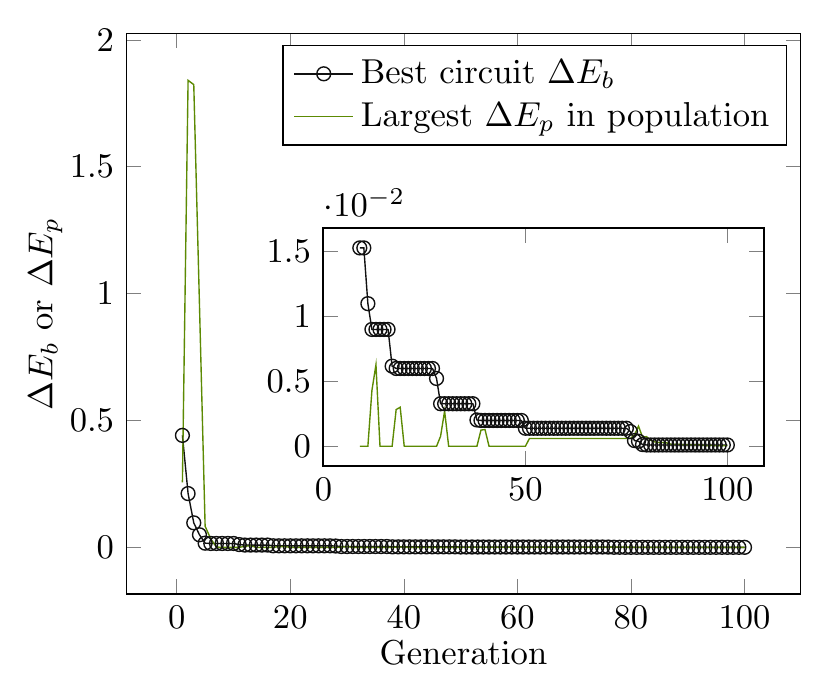}
	\caption{\label{fig:EnergyConvergence} The plus symbols indicate the energy difference between the ground state
		and that yielded by the best individual at that generation; the open circles indicate the largest energy
		difference within the population at that generation. The inset shows a magnification of the figure
		starting at generation eight.}
\end{figure}

Figure~\ref{fig:EnergyConvergence} shows the convergence of \codeName{QuantumGEP} to a perfect circuit: The
x-axis is the generation and the y-axis is (i) for the plot indicated by plus symbols, the energy difference $\Delta E_b$ between the ground state
and that yielded by the best individual of that generation, and (ii) for the plot indicated by open circles, the largest energy
difference $\Delta E_p$ within the population of that generation. The first generations show large improvements in
quantum circuit quality (open circles), and circuit quality continues to improve but with smallest changes in $\Delta E_b$
for late generations. In addition, variations in the population of circuits (solid green line) show a large diversity increase for the first generations. Even though later these population variations are much smaller due to evolutionary pressures, the circuits themselves are all different: they just yield either the same energy or an energy very close to the exact one. 

\subsection{Quantum Chemistry Applications}

\codeName{QuantumGEP} can find a quantum circuit, which yields the ground state of a quantum chemistry problem, when applied to an initial state.
We have considered only basic quantum gates, and have applied \codeName{QuantumGEP} to two molecules: NaH and the benzene molecule.

A four bit Hilbert space models the NaH molecule, and the ground state resides in the two electron sector; see \cite{re:mccaskey19} for the full quantum computing solution.
One basis state accounts for over 99\% of the ground state weight: the Hartree Fock solution; three other
basis states contribute the rest with a combined weight of less that 1\%. \codeName{QuantumGEP} easily yields circuits for the Hartree Fock state. For
example, Ry$_1:\pi$ P$_2$ Ry$_3:\pi$ P$_3$ with fitness 160.29(9)), which equals the opposite of the Hartree-Fock energy, whereas
the true ground state energy is -160.30137813(6) Hartrees. 
In our tests, \codeName{QuantumGEP} was unable to
produce the other basis states that contribute to the ground state, as we discuss in section~\ref{sec:failures}.

An eight bit state models the benzene molecule, with the ground state in the four electron sector. Once more, the Hartree-Fock state
accounts for most of the weight, this time at over 95\%.
\codeName{QuantumGEP} easily yields circuits that produce the Hartree Fock solution, such as
Ry$_0:\pi$ P$_0$ Ry$_1:\pi$ P$_2$ Ry$_4:\pi$ Ry$_5:\pi$ P$_7$ with fitness 8.89(4) or the opposite of the
Hartree-Fock energy. By decreasing the maximum circuit size or \keyword{Headsize} in the input file, even smaller circuits
can quickly be found, such as Ry$_0:\pi$ Ry$_1:\pi$ Ry$_4:\pi$ Ry$_5:\pi$.
The other contributions to the ground state are more difficult to find, and require
enabling conjugate gradient from the input,
as well as a rescaling of the energy in the input:
$H'=10(H-E_{HF}),$ where $E_{HF}$ is the Hartree-Fock energy, or values close to it.
Obtaining all these small contributions to the ground state will be the focus of future work.

\section{Convergence and Acceleration}
\subsection{Convergence Failures} \label{sec:failures}
We have observed a failure of convergence in the case of the NaH system for four bits\cite{re:mccaskey19}, where the ground
state equals $-0.99854285211(3)|1010\rangle+\cdots$, where the ellipsis indicates states with weight to complete the normalization.
We surmise that the failure occurs due to the very high weight of a single basis state, and would require for its solution the
use of modular gates so that a more sophisticated quantum circuit can be constructed by the artificial intelligence.

We have also found a convergence failure in the two-dimensional quantum Heisenberg model on a 3$\times$3 lattice or 9 bits.
In this case, we believe the lack of convergence results from the large number of one- and two-bit gates that would be required
to express the ground state of the Heisenberg model in this case. We know that the solution can be written in shorter form
by using many bit gates in a multi-scale entanglement answer representation \cite{PhysRevLett.101.110501}. Here again we think the
problem would be solved by modular gates.

\subsection{Shared Memory Parallelization}
\codeName{QuantumGEP} presents multiple opportunities for shared memory parallelization, and we have implemented two of them.
The need to apply a matrix to a vector leads to the use of \codeName{BLAS}'s \cite{blackford2002updated} \codeName{ZGEMM} call that can be parallelized over
multiple cores on done on the GPU; these matrix-vector operations appear in Hamiltonian actions or gate actions.
Moreover, the fitness of each quantum circuit of a given generation can be calculated in parallel, and we have implemented this
in pure C++ using PsimagLite's Parallelizer2, which supports either an openMP or a \emph{pthreads} back-end.

\subsection{Source Code and Documentation}
We have released both the scientific application \codeName{QuantumGEP} and its computational engine \codeName{evendim}
at \nolinkurl{https://code.ornl.gov/gonzalo_3/evendim} and at
\nolinkurl{https://github.com/g1257/evendim},
with a free and open source license. They are written fully in \textsc{C++}; the documentation, test suite location, dependencies,
and other details
are given in the supplemental material \todo{URL will be given by publisher}.

\section{Summary and Outlook}\label{sec:summary}
Quantum circuits have been created manually, and only recently has artificial intelligence been used in this field \cite{https://doi.org/10.48550/arxiv.2110.07441}.
An automatic tool can come to the aid of quantum circuit generation, even
when manual intuition remains important. 
This paper has then introduced \codeName{QuantumGEP}, an automatic generator of quantum circuits to produce
the ground state of quantum and classical Hamiltonians, based on gene expression programming.
\codeName{QuantumGEP} is free and open source, with a computational engine (\codeName{evendim})
agnostic to fitness function, to primitives (or gates) and to leaves (or inputs), so that problem specific fitness and primitives
can be written without modifying \codeName{evendim}.

The richness of GEP \cite{re:ferreira06} allows for constants, automatically defined functions, and other features
included in \codeName{evendim}; GEP's use of ``junk DNA'' allows for almost unrestricted mutations, improving convergence.
\codeName{QuantumGEP} was successfully applied to the generation of circuits for the MaxCut problem, and it was able
to yield intuitive answers in simple condensed matter problems. In quantum chemistry it successfully yielded the
Hartree-Fock ground eigenstate, but more work remains to obtain more precise quantum circuits for the exact ground state.
We think this will entail the use of modular gates, which can be built as combination of basic gates, where automatically
defined functions (ADF) and multiple ``genes'' would improve convergence. Although ADFs and multiple genes are already implemented
in \codeName{evendim}, we did not want to expand its application to difficult quantum chemistry problems in this introductory paper, and
leave those for future study.

\section*{Acknowledgments}
This work was performed at Oak Ridge National Laboratory, operated by UT-Battelle, LLC under contract DE-AC05-00OR22725 for the US Department of Energy (DOE). Support for the work came from the DOE Advanced Scientific Computing Research (ASCR) Accelerated Research in Quantum Computing (ARQC) Program under field work proposal ERKJ354.

\input{paper108.bbl}

\end{document}